\documentclass[12pt,preprint]{aastex}

\begin{document}

\title{The X-ray Asynchronous Optical Afterglow of GRB 060912A and
    Tentative Evidence of a 2175-\AA\, Host Dust Extinction Feature}

\author{J.~Deng\altaffilmark{1},
 W.~Zheng\altaffilmark{1,*},
 M.~Zhai\altaffilmark{1},
 L.~Xin\altaffilmark{1},
 Y.~Qiu\altaffilmark{1},
 A.~Stefanescu\altaffilmark{2,**},
 A.~Pozanenko\altaffilmark{3},
 M.~Ibrahimov\altaffilmark{4},
 A.~Volnova\altaffilmark{5}
 }

\altaffiltext{1}{National Astronomical Observatories, 20A Datun Road, Chaoyang District, Beijing 100012, China;
jsdeng@bao.ac.cn}

\altaffiltext{2}{Max-Planck-Institut f\"{u}r Extraterrestrische Physik, Giessenbachstrabe, 85740 Garching, Germany}

\altaffiltext{3}{Space Research Institute of RAS, Profsoyuznaya, 84/32, Moscow 117997, Russia}

\altaffiltext{4}{Ulugh Beg Astronomical Institute, Tashkent 700052, Uzbekistan}

\altaffiltext{5}{Sternberg Astronomical Institute, Moscow State University, Moscow 119992, Russia}

\altaffiltext{*}{now at Department of Physics, University of Michigan, Ann Arbor, MI 48109, USA}

\altaffiltext{**}{now at Max-Planck-Institut Halbleiterlabor, Otto-Hahn-Ring 6, 81739 M\"{u}nchen, Germany
and Johannes Gutenberg-Universit\"{a}t, Inst. f. anorganische und analytische Chemie, 55099 Mainz, Germany}

\begin{abstract}

We present optical photometry of the GRB 060912A afterglow obtained with ground-based telescopes,
from about 100~sec after the GRB trigger till about 0.3~day later, supplemented with the $Swift$
optical afterglow data released in its official website. The optical light curve (LC) displays
a smooth single power-law decay throughout the observed epochs, with a power-law index of about
-1 and no significant color evolution. This is in contrast to the X-ray LC which has a plateau
phase between two normal power-law decays of a respective index of about -1 and -1.2. It is shown
by our combined X-ray and optical data analysis that this asynchronous behavior is difficult
to be reconciled with the standard afterglow theory and energy injection hypothesis.
We also construct an optical-to-X-ray spectral energy distribution at about 700~sec after
the GRB trigger. It displays a significant flux depression in the $B$-band, reminding us of
the possibility of a host-galaxy (at $z=0.937$) 2175-\AA\ dust absorption similar to
the one that characterizes the Milky Way extinction law. Such an identification, although
being tentative, may be confirmed by our detailed analysis using both template extinction
laws and the afterglow theory. So far the feature is reported in very few GRB afterglows.
Most seem to have a host galaxy either unusually bright for a GRB, just like this one,
or of an early type, supporting the general suggestion of an anti-correlation between the
feature and star-forming activities.

\end{abstract}

\maketitle

\section{Introduction} \label{sec-intr}

The launch of $Swift$ satellite in late 2004 has brought about a new era in the field of gamma-ray bursts (GRBs).
More than 95\% of the $\sim 100$ $Swift$ GRBs reported each year had their afterglows detected in X-rays
and about 60\% detected in the optical, which in turn led to a large fraction of $Swift$ GRBs having redshift
determination and detailed studies \citep{geh09}. New questions have also arisen as the cases of well-observed
afterglows keep accumulating, which pose new challenges to the standard GRB afterglow model (for recent reviews
see \citealt{zha07a,zha07b,gra08,pan08b,geh09}). One of these puzzles is the occasional lack of an optical light
curve (LC) behavior synchronous to a LC break of the X-ray afterglow that usually indicates the end of a shallow
decay phase (e.g., \citealt{fan06,pan06,mel08}; but see \citealt{oat09}).

The incredible brightness of GRBs makes them excellent probes of the distant universe, e.g., to
explore the dust extinction properties of their host environments. Typical attempts involve fitting
the afterglow spectral energy distributions (SEDs) with template extinction laws, although a more
sophisticated approach based on a parameterized model has been proposed \citep{lia08}. Most studies
favor a Small Magellanic Cloud (SMC) extinction law \citep{str04,kan06,sta07,sch07}. However,
the broad 2175-\AA\ bump characteristic of the Milky Way (MW) extinction law (also marginally appearing
in that of the Large Magellanic Cloud (LMC)), whose physical carrier remains a long-standing mystery
\citep{dra03}, was recently identified in a small number of GRBs \citep{kru08,eli09,pro09,lia09}.

In this paper, we present the observations of the $Swift$ GRB 060912A optical afterglow, which shows a
LC behavior inconsistent with that of the X-ray one, and possible evidence of a 2175-\AA\ host dust
extinction feature in the SED. The gamma-ray and X-ray properties of the GRB are summarized in \S \ref{sec-he}.
Our ground-based observations and the telescopes used, as well as the $Swift$ Ultra-Violet and Optical
Telescope (UVOT) data that are extracted from the official online catalogue, are described in \S \ref{sec-oa}.
Combined analysis of the optical and X-ray data are detailed in \S \ref{sec-pla}, within the context of
the standard afterglow model and energy injection hypothesis while also revealing discrepancies to the models.
An optical-to-X-ray SED is built in \S \ref{sec-sed}, and the presence of the 2175-\AA\ feature is argued
for through extinction template fitting after an simplified correction for H Lyman absorption.
A discussion on the implications of our results is given in \S \ref{sec-dis}.

\section{Gamma-Ray and X-Ray Characteristics} \label{sec-he}

GRB 060912A triggered the $Swift$ Burst Alert Telescope (BAT) and Konus-Wind \citep{hur06a,gol06}.
Its single-pulse LC lasted for a duration of $T_{90}\sim6-8$~sec, and hence a long burst as confirmed
by spectral lag analysis \citep{par06} and by its star-forming host galaxy at $z=0.937$ \citep{lev07}.
The BAT data were fitted with a power-law spectrum of a photon index $\Gamma_{\rm BAT}\approx -1.74$ and
a fluence of $\sim 1.35\times 10^{-6}$~erg~cm$^{-2}$ \citep{sak08}. The rest-frame $E_{\rm peak}$ may be
$\sim 150$~keV according to \citet{lia07} (based on an empirical $E^{\rm obs}_{\rm peak}-\Gamma_{\rm BAT}$
correlation; see also \citealt{sak09}). Assuming a high-energy index of between $-2$ and $-3$,
the isotropic luminosity is $\sim 0.7-1\times 10^{52}$~erg, satisfying the Amati relationship
within $2\sigma$ (e.g., \citealt{ama08})

\citet{lia07} re-analyzed the X-ray afterglow data of the GRB as observed with the $Swift$ X-Ray
Telescope (XRT), independent of the preliminary results of \citet{hur06b}. For the time-integrated
XRT spectrum ($[0.3-10$~keV]), they obtained a photon index of $\Gamma^{\rm mid}_{\rm X}=-2.08\pm0.11$
before $\sim 1,350$~sec and of $\Gamma^{\rm late}_{\rm X}=-1.95\pm0.13$ after that, correcting for
absorption by both the Galaxy and the host galaxy. The fitting H-column density assuming a solar
metallicity was $N^{\rm host}_{\rm H}\approx 4.2\times10^{21}$~cm$^{-2}$ at $z=0.937$ for the host
galaxy and about $3.8\times 10^{20}$~cm$^{-2}$ for the Galactic value in the GRB direction.

We reproduce in Figure 1 ({\em open circles}) the XRT LC as reduced by those authors, which consists
of an early deep decay before $\sim 350$~sec, a late deep decay after $\sim 1,800$~sec, and a plateau
in between. The three LC segments can be fitted using power law $f_{\rm X}\propto t^\alpha$ separately,
with the power-law indices being $\alpha^{\rm early}_{\rm X}=-1.04\pm0.16$, $\alpha^{\rm mid}_{\rm X}=-0.19\pm0.08$,
and $\alpha^{\rm late}_{\rm X}=-1.19\pm0.05$, which is consistent with the smoothed broken power-law
model results in that paper.

\section{Optical Afterglow Observations} \label{sec-oa}

\subsection{Xinglong Observations} \label{sec-xl}

The 0.8-m Tsinghua-NAOC Telescope (TNT) at the Xinglong Observatory automatically observed the GRB
upon receiving the GCN alert. The telescope is equipped with standard Johnson-Cousin $UBVRI$ filters
and a Princeton Instrument $1340\times1300$ CCD, resulting a field of view of $\sim 11'\times11'$.
For the details of the Xinglong GRB follow-up system, see \citet{zhe08} and \citet{den06,den08}.

The TNT observation started $\sim24$~sec earlier than the UVOT and lasted for more than 3.7~hr. It began
with a preset sequence of unfiltered exposures, i.e., in a $white$ band, and was later switched to
the filtered bands. Extra observations were made with a 60-cm telescope. The afterglow was detected in
the $white$ and $R$ bands, but not in the $B$ and $V$ bands.

The observation results are tabulated in Table 1. The data were reduced using IRAF in a standard
routine, bias subtracted and flat-fielded. Differential PSF photometry was performed using 8 reference stars,
which were calibrated in a photometric night. Since the response curve of our CCD peaks in the $R$ band,
we used the $R$-band magnitudes of the reference stars for the $white$ band and assumed that the so-derived
afterglow magnitudes are equivalent to $R$-band ones. Our independent photometric tests with stellar objects
have shown this to be a decent assumption, introducing a typical scatter of only $\sim 0.07$ mag.

\subsection{Maidanak Observations} \label{sec-mai}

The 1.5-m telescope of Maidanak Observatory is located at the south-east of Uzbekistan. Its follow-up
observations were made in both the $R$ and $B$ bands from about 3 to 6 hours after the GRB onset under
favorable weather and seeing of $\sim 0.6''$, using an SITe $2000\times800$ CCD with a field of view
of $\sim 8.5'\times3.5'$. The data were reduced in the same way as the Xinglong data, using the same
set of reference stars and photometric calibrations. The results are also listed in Table 1.

\subsection{Skinakas Observations} \label{sec-opt}

About 5.5~hr after the GRB, the afterglow was also caught by the Roper/Photometrics CH360 CCD camera
(SITe $1024\times1024$ chip), which was mounted in parallel with the OPTIMA-Burst instrument on
the 1.3-m Skinakas Observatory telescope at Crete, Greece \citep{ste07}. The field of view of
the CCD camera is $\sim 8.5'\times8.5'$. Again the data were reduced in the same way as the Xinglong
data and included in Table 1.

\subsection{Swift UVOT Observations} \label{sec-uvot}

The $Swift$ UVOT photometry in Table 1 was derived using data extracted from the official online UVOT GRB
afterglow catalogue \citep{rom09}. The UVOT started observing the GRB since $\sim 110$~sec after the BAT trigger.
A fading afterglow was detected in the $UVOT$-$white$, $V$, $B$, $U$, $UVW1$, and $UVM2$ bands, but not
in the bluest $UVW2$ band. When needed we combined raw data of neighboring epochs, taking both the coincidence
loss corrected source count rate and background count rate, to make signal-to-noise ratio (S/N) larger than 3.
But in the $B$, $UVW1$, and $UVM2$ bands we had to keep the data with $2.6\lesssim{\rm S/N}\lesssim 3$ which
could mean a marginal detection.

The $UVOT$ observations lasted till very late time, $\sim 10$ days in the $U$ band, $\sim 3.4$ days
in the $UVOT$-$white$ band, and $\sim 1.4$ day in the rest. From the very late-time detections, which are
not included in Table 1, we derived host-galaxy brightness of about $21.25\pm 0.15$, $21.75\pm 0.05$,
and $21.5\pm 0.4$~mag in the $U$, $UVOT$-$white$, and $UVW1$ bands, respectively.

\subsection{Multi-band Optical Light Curves} \label{sec-opt}

As shown in Figure 1, the flux of the optical afterglow decayed smoothly following a single power law
in all bands and throughout the whole observation period, including the X-ray plateau phase. This is
best demonstrated by the well-sampled $R$-band LC. The fitting power-law indices $\alpha$ of the flux evolution
$f\propto t^\alpha$ are all similar, i.e., $-1.05\pm0.02$ for $R$, $-0.94\pm0.11$ for $V$, $-0.98\pm0.06$
for $B$, $-0.86\pm0.07$ for $U$, $\sim -1$ for $UVW1$, and $-0.90\pm0.18$ for $UVM2$. The $R$-band LC
was built using both the genuine $R$-band observations and the unfiltered TNT observations which have
been approximately reduced to the $R$-band. But the Maidanak and Skinakas $R$-band data after 10,000~sec
were not included in the fitting process due to the relatively large scatter among the data.

The late-time photometry must have been contaminated to some extent by the light of the host galaxy,
which has been found by \citet{lev07} to be one of the brightest for GRBs. Those authors measured
$R=22.0\pm0.5$~mag, while using the host-galaxy spectrum published by them we derived $V\sim 22.4\pm0.5$~mag.

The optical LCs after subtracting the host galaxy contribution can also be well fitted by a single
power law, with the slopes of each band slightly steeper but closer to one another, than before
the subtraction. Specifically, we obtained $\alpha_R=-1.08\pm 0.02$, $\alpha_V=-0.98\pm 0.11$,
$\alpha_U=-1.10\pm 0.10$, and $\alpha_{UVW1}\sim -1.1$. The values are also similar to those of
the X-ray LC before and after the plateau phase.

The optical afterglow did not evolve in synchronism with X-rays, being lack of any LC features that would
correspond to the two breaks defining the beginning and end of the X-ray plateau. Similar behaviors were also
observed in a few other bursts detected in the $Swift$ era, e.g., GRB 050319 \citep{qui06} and GRB 070420
\citep{klo08}. 
There have been many theoretical discussions (e.g., \citealt{fan06,pan06}) on that deviation to the standard
afterglow model, but no satisfactory interpretation so far.

\section{The X-ray Plateau, Fast Optical Decay and Energy Injection Model} \label{sec-pla}

It has become well known that the X-ray plateau phase can not be understood with the standard afterglow model
\citep{fan06,pan06}. The standard model gives a shallow LC decay only if $\nu_c<\nu<\nu_m$, where $\nu_m$ is
the minimum synchrotron frequency and $\nu_c$ the cooling frequency (see \citealt{zha04}). The model decay
index $\alpha=-0.25$ may be similar to $\alpha^{\rm mid}_{\rm X}=-0.19\pm0.08$ of GRB 060912A, but the model
photon index $\Gamma=-1.5$ is clearly inconsistent with the observed $\Gamma^{\rm mid}_{\rm X}=-2.08\pm0.11$.

During the X-ray plateau phase of the GRB, the X-ray frequency must be $\nu_{\rm X}>\max\{\nu_m,\nu_c\}$,
while the optical one $\nu_{\rm opt}$ is probably between $\nu_m$ and $\nu_c$, in the context of both the X-ray
and optical afterglows coming from the same forward-shock synchrotron radiation process. The model spectral
index $\beta\equiv\Gamma+1$ as a function of the electron index $p$ is $-p/2$ for $\nu>\max\{\nu_m,\nu_c\}$,
$1/3$ for $\nu<\min\{\nu_m,\nu_c\}$, and in between either $(1-p)/2$ (slow cooling) or $-1/2$ (fast cooling).
The expressions are the same between the standard model and its extension to include the energy injection
scenario or that for the $1<p<2$ case (see \citealt{zha04} and \citealt{zha06}). Our analysis of the spectral
energy distribution (SED) in the following section indicates that the intrinsic optical spectral index is
probably $>-0.7$. Therefore the optical photons must come from a model spectral segment different from that of
the X-rays. On the other hand, $\nu_{\rm opt}$ can not be $<\min\{\nu_m,\nu_c\}$ since it is highly unlikely
that the extinction was so large as to make a positive intrinsic $\beta_{\rm opt}$. Finally, to reconcile
the spectral index $(1-p)/2$ between $\nu_m$ and $\nu_c$ with the observed $\beta^{\rm mid}_{\rm X}=-1.1$
would require an electron spectrum of $p\gtrsim3$ which is too steep.

The X-ray plateau of the GRB may be explained using the energy injection model, which assumes an energy injection
rate proportional to $t^{-q}$ ($q<1$) \citep{zha06}. Since $\nu^{\rm mid}_{\rm X}>\max\{\nu_m,\nu_c\}$, we obtain
$p\approx 2.2\pm0.2$ by equating $-p/2$ with $\beta^{\rm mid}_{\rm X}\approx -1.1\pm0.1$, which is most reasonable
for an astronomical relativistic electron spectrum. The model predicts that $p=-2+(8-4\alpha^{\rm mid}_{\rm X})/(q+2)$,
and hence $q\approx 0.07\pm0.12$. Such a small $q$ seems close to the case of energy injection due to pulsar spin-down
\citep{dai98,zha01}, and is consistent with the average value of $-0.07\pm 0.35$ found by \citet{lia07} for their
$Swift$ sample.

The X-ray LC and spectrum after the plateau phase as the energy injection has ended also suggests an electron index
of $\sim 2.1-2.2$. The observed $\alpha^{\rm late}_{\rm X}=-1.19\pm0.05$ and $\Gamma^{\rm late}_{\rm X}=-1.95\pm0.13$
satisfy the closure relationship $\alpha(\beta)$ only if $\nu>\max\{\nu_m,\nu_c\}$, in which case the standard model
has $\alpha=(2-3p)/4$ and $\beta=-p/2$ \citep{zha04}.

However, the energy injection hypothesis is not applicable to the simultaneous fast optical LC decay observed throughout
the X-ray plateau phase. Only for the slow-cooling case of the wind scenario, the model LC between $\nu_m$ and $\nu_c$
can decay faster than $\nu>\max\{\nu_m,\nu_c\}$. But the model prediction, $\alpha_{\rm opt}=\alpha_{\rm X}-(2-q)/4$,
would compare with the observation, i.e., $\alpha_{\rm opt}\sim-1$, only if $q\lesssim -1$, contradicting the above
estimate of $q\approx 0.07\pm0.12$ as constrained by the observed $\beta^{\rm mid}_{\rm X}$ and $\alpha^{\rm mid}_{\rm X}$.

\section{The Spectral Energy Distribution and Dust Extinction} \label{sec-sed}

We select $t=700$~sec to build the optical-to-X-ray spectral energy distribution (SED), shown in Figure 2 as
{\em filled squares}. Around the epoch, observational data can be found in the largest number of photometric
bands. The host-subtracted magnitudes are listed in Table 2. The small flux change between the real data epoch
and $t=700$~sec has been corrected for, adopting the LC power-law indices as given in \S \ref{sec-opt},
as well as a small Galactic reddening of $E(B-V)=0.05$. For the $B$ and $UVM2$ bands of no host measurement,
we assumed that the intrinsic afterglow flux should be $\sim 0.05$~mag fainter than the total, taking the
host-subtracted results in the other bands as references. The optical magnitudes were converted into monochromatic
fluxes using the conversion factors and effective frequencies taken from \citet{poo08} and \citet{bes98}.
The X-ray flux is drawn at 1.73~keV in the figure, with {\em dotted lines} indicating spectrum uncertainties.

For comparison, we built also the SED at $t=7,000$~sec (Figure 2; {\em open squares}). Only the optical data
in the $R$, $V$ and $U$ bands and the X-ray data were used in order to avoid large uncertainties caused by
host galaxy contamination and data interpolation.

By power-law fitting, we obtained an optical-to-X-ray spectral index of $\beta_{\rm OX}=0.7\pm0.1$ for both epochs.
First, the GRB clearly does not belong to the ``optically'' dark bursts ($\beta_{\rm OX}<0.5$; \citealt{jak04,zhe09}).
Secondly, at $t=700$~sec the optical fluxes in the $UVW1$ and $UVM2$ bands fall much below the optical-to-X-ray
fitting power law, suggesting either strong dust extinction in the host galaxy or hydrogen Lyman line absorption.
We further suggest that the intrinsic spectrum of the optical afterglow during the X-ray plateau phase is probably
shallower than $\nu^{-0.7}$ and hence the optical band lies between $\nu_m$ and $\nu_c$ at the epoch. If we instead
assume an intrinsic optical spectral index equivalent to the X-ray one, the extinction-corrected $\beta_{\rm OX}$
will still be considerably larger than $\beta_{\rm X}^{\rm mid}=-1.08$ for any of the MW ($\beta_{\rm OX}\approx -0.86$),
LMC ($\beta_{\rm OX}\approx -0.78$), and SMC ($\beta_{\rm OX}\approx-0.75$) extinction laws. A special artificially
fine-tuned flat extinction law would be required to make the optical-to-X-ray SED as steep as the X-ray spectrum
itself, which is beyond the scope of this paper (but see \citealt{che06}).

The flux loss caused by H Lyman absorption was $\sim 0.11$~mag in the $UVM2$ band and $\sim 0.07$~mag in the $UVW1$
band. Most absorption probably occurred in the host-galaxy damped-Ly$\alpha$ system that gave rise to the large
X-ray $N^{\rm host}_{\rm H}$ of $\sim 4.2\times10^{21}$~cm$^{-2}$, as estimated by us assuming a Voigt line profile
taking into account thermal-Doppler broadening and natural broadening. The intergalactic hydrogen contributed only
$\sim 0.02$~mag in $UVM2$ and $\sim 0.01$~mag in $UVW1$ at $z=0.937$, according to the formulae of \citet{mad96}.
In our calculations, when doing flux integration, a power law spectrum of $f_\nu\propto \nu^{-0.6}$ was assumed
and the $UVOT$ filter response curves of \citet{poo08} were adopted.

In the 700-sec SED, the $B$-band flux falls well below both the $V$ band flux and the $U$ band one, which seems
to suggest the interesting possibility of a 2175-\AA\, dust extinction feature from the host galaxy. One could
argue that the $V$-band flux may be overestimated since it looks larger than the $R$-band one. But we note that
it is also above the simple interpolation from the $R$-band to the $U$-band, so evidence of extra extinction
in the $B$-band is unambiguous. Such a feature is characteristic of the dust extinction law of the MW, displaying
a full width of $\sim 700$~\AA\, \citep{car89}. It does not exist in the extinction law of the SMC, while being
rather modest in the LMC one \citep{pei92}. At $z=0.937$, it is almost entirely located in the observer-frame
$UVOT$ $B$-band \citep{poo08}.

We tested the three dust-extinction laws on the 700-sec optical SED, with the results shown in the $inset$ of
Figure 2. Note that the $UVW1$ and $UVM2$ bands have been corrected for the H Lyman absorption, after which
the optical SED ({\em filled squares}) can be fitted with a power law of index $\sim-1.7\pm0.4$. As discussed
above, the afterglow model predicts an intrinsic spectral index $\beta_{\rm opt}$, i.e., after correction for
dust extinction, of either $-0.5$ (fast cooling) or $(1-p)/2$ (slow cooling), i.e., $\sim -0.5 - -0.7$ assuming
$p\sim 2.0-2.4$. Thereafter, we took $-0.6\pm0.1$ as the reference value of $\beta_{\rm opt}$ and reproduced
it with the extinction-corrected optical SED by adjusting the reddening value $E(B-V)$. The results are
$E(B-V)=0.25\pm0.02$ for the MW extinction law, $\sim 0.15$ for the LMC case, and $\sim 0.1$ the SMC.

As shown in the figure ({\em filled circles}), the MW-like extinction well corrects the observed flux deficiency
in the $B$ band. This is just as expected and in stark contrast to the cases of LMC ({\em open circles})
and SMC ({\em open suares}). Note that here we did not use the common minimum-$\chi^2$ approach to constrain
the extinction. A minimum power-law fitting $\chi^2$ were obtained for a much smaller $E(B-V)$, resulting in
a corrected optical SED too steep to comply with the general afterglow model and a reasonable extension to the X-ray
spectrum.

A MW-like dust extinction law may also be supported by the fact that the host galaxy is one of the brightest
among long GRBs (\citealt{lev07}; see also \citealt{sav09}). Brighter galaxies tend to have an evolved stellar
population and higher metallicity. After correcting the published host galaxy spectrum for redshift, we synthesized
its absolute $B$-band magnitude and obtained $M_B<-21.5$ even before extinction corrections. So the galaxy
must have metallicity near or even higher than the solar value, or $Z\equiv 12+\log({\rm O}/{\rm H})>8.9\pm0.6$
according to the \citet{kob04} $M_B$-$Z$ relationship for $z\in[0.8,1.0]$. It is significantly higher than most
GRB host galaxies ($\sim 7.4-8.7$; \citealt{sav09}), LMC ($\sim 8.3$), and SMC ($\sim 8.0$). On the other hand,
the dust-to-gas ratio, $A_V/N^{\rm host}_{\rm H}$, is smaller than the empirical values of the MW, LMC, and SMC,
just as the other GRB host galaxies \citep{sch07}.

Although the GRB is not optically dark, it may belong to the ``dim'' category of optical afterglows as defined
by \citet{lia06} and \citet{nar06}. Using the two fitting LCs shown in Figure 1, we derived the
host-subtracted observer-frame $R$-band magnitude at $(1+z)12$hr of $\sim 23.1\pm 0.4$, corrected for the
Galactic extinction. This was then converted to a rest-frame 12-hr $R$-band luminosity assuming the observed
optical SED of $f_{\nu}\propto \nu^{-1.7\pm0.4}$ (see Figure 2). Finally, after the host-galaxy extinction
of $A_R\sim 0.5$ (MW-like law) was corrected for, we obtained an intrinsic afterglow luminosity of
$L(\nu_R,12{\rm hr}) \sim 1.9\pm 0.8\times 10^{29}$~erg. The latest dim afterglow sample, compiled by
\citet{nar08}, has a mean value of $\langle\log L(\nu_R,12{\rm hr})\rangle=29.3$.

\section{Discussion} \label{sec-dis}

With GRB 060912A, we have added another case to a few puzzling GRBs that show achromatic X-ray LC breaks
at the beginning and end of a plateau phase which are not present in the optical band. Unlike the
original sample compiled by \citet{pan06} and those supplemented by other authors (e.g., \citealt{mel08}),
in current case the optical temporal index is quite similar to the contemporaneous X-ray one before
and after the X-ray plateau. Although the optical and X-ray characteristics during the X-ray plateau
are difficult to explain self-consistently just as the others, the standard afterglow model can actually
apply to the rest stages of this GRB. The optical afterglow showed little if any evolution in its temporal
index and optical colors, so it likely arose from the same synchrotron spectral segment throughout,
or specifically $\nu_m<\nu_{\rm opt}<\nu_c$ of the ISM and slow-cooling case, while the post-plateau
X-ray afterglow must have $\nu_{\rm X}>\nu_c$ as demonstrated by its slightly steeper LC slope and
a significant spectral deviation from the optical-to-X-ray SED. There could be a cooling break around
the beginning of the X-ray plateau, but we do not have the early X-ray spectrum to either argue for
or against this. In overall our data analysis seems to support a hypothesis that the optical and X-ray
afterglows are of the same origin except for the X-ray plateau which is dominated by an extra component
\citep{pan08a}. Other explanations like a reverse-shock afterglow \citep{uhm07,gen07} and the ``late prompt''
emission \citep{ghi09} are not favored for this GRB.  An evolution of microphysical parameters
(e.g., \citealt{fan06,iok06}) is not excluded providing that it is strictly associated with the energy
injection process. A quantitative comparison between our data and the theory of \citet{pan08a} may be
possible but is beyond the scope of this paper.

We have shown evidence of a 2175-\AA\ extinction feature in the optical SED of the GRB 060912A
afterglow. Such a feature has only been observed in a very small number of GRBs, which may also
include GRB~970508 \citep{str04}, GRB~991216 \citep{kan06,vre06}, GRB~050802 \citep{sch07},
GRB~070802 \citep{kru08,eli09}, GRB~080607 \citep{pro09}, and possibly even the high-redshift
GRB~050904 \citep{lia09}. Only the identification in GRB~070802 can be regarded as robust since
the feature was detected not only through multi-color photometry but also clearly in a spectrum.
In the other cases, the feature was merely revealed as a dip in the corresponding passband
in the observed optical SED. Although the afterglow spectrum of GRB~991216 also displayed
a similar feature, it centers at a somewhat ``wrong'' position, about 2360 \AA, \citep{vre06}.
This UV absorption bump is the most prominent feature of the MW extinction law and is now widely
hypothesized to be caused by a mixture of large polycyclic aromatic hydrocarbon molecules or clusters,
whose strong C-C bond $\pi\rightarrow\pi^{*}$ transitions peak at $\sim 2175$~\AA, and small
carbon-rich grains (\citealt{dra03} and references therein). Both species are thought to be fragile
under the illumination of strong UV and X-ray radiation (e.g., \citealt{voi92,fru01}), so the reported
absorption likely took place a few parsecs away from the GRB location (\citealt{per02}; but see
\citealt{lia09}). It was suggested that the UV bump requires a relatively evolved stellar population,
which means less dust/molecule destruction by UV photons or interstellar shocks, in order to explain
its absence in the SMC and most starburst galaxies \citep{gor97,nol07}. So it is not surprising that
the majority of GRB afterglow SEDs are best fitted using a SMC extinction law (e.g., \citealt{kan06,sch07,sta07})
since they are typically hosted in faint dwarf star-forming galaxies. On the other hand, among the few cases
with a reported UV bump, GRB~050802, GRB~060912A, and GRB~070802 all have a host galaxy unusually
bright (possibly $M_B\lesssim -21--22$; see \citealt{fyn05}, \citealt{eli09}, \citealt{lev07},
and this paper) for a GRB, while the faint host galaxy of GRB~970508 has an early-type morphology
\citep{sav09}. Finally we note that when the paper is written a statistical study made by
\citet{sch09} is published which also presents the UVOT SED of GRB 060912A, but those authors
seem prudent not to claim the observed $B$-band flux depression a detection of the 2175-\AA\ feature.

\acknowledgements

We thank E. Liang for the electronic files of the X-ray data. This work was supported
by the National Basic Research Program of China (Grant No. 2009CB824800) and by the NSFC
(Grant No. 10673014).

\newpage

\plotone{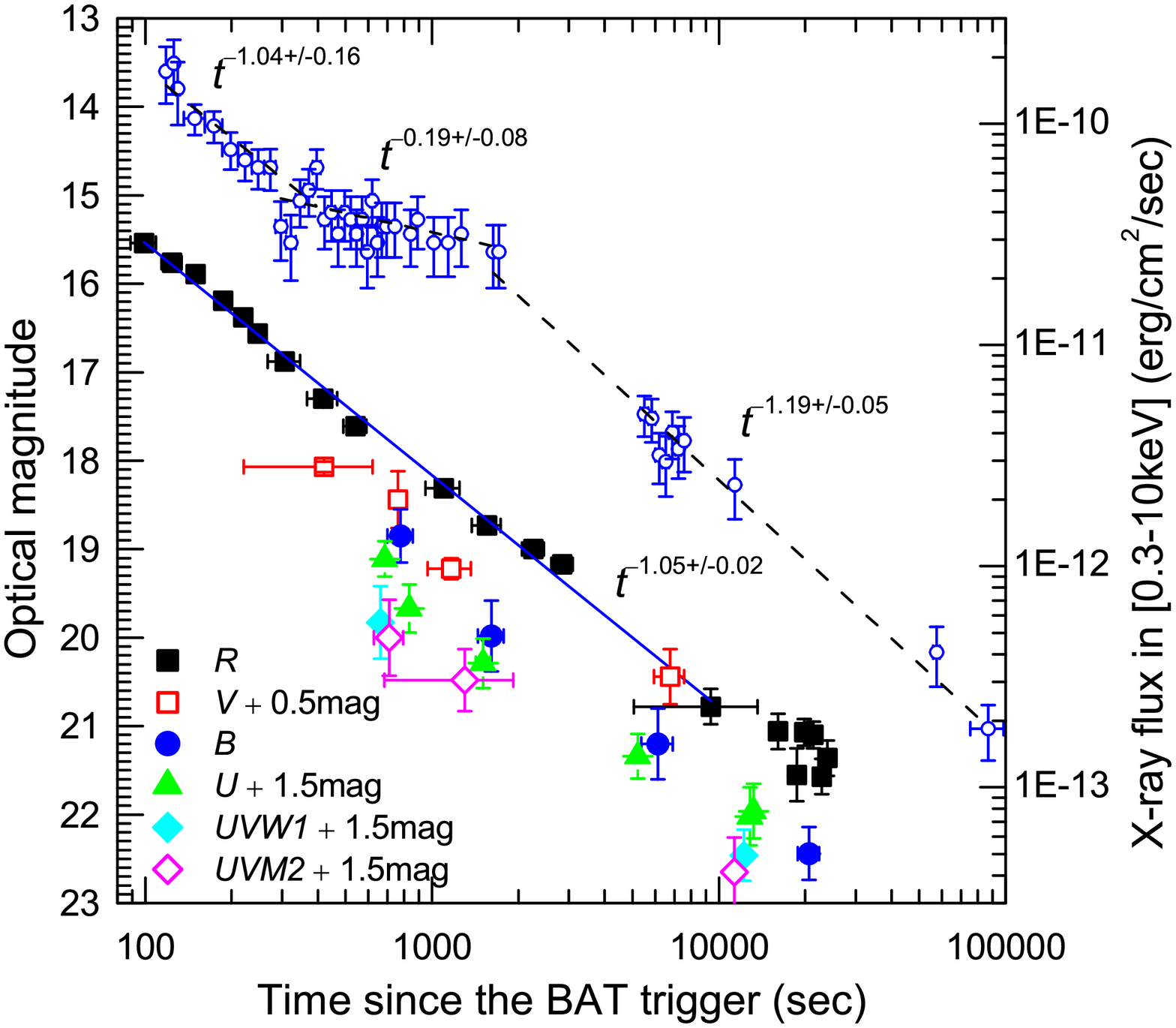}

\figcaption[f1.eps]{Observed optical light curves of GRB 060912A afterglow plotted in magnitude
in the $R$ ({\em filled squares}), $V$ ({\em open squares}), $B$ ({\em filled circles}),
$U$ ({\em triangles}), $UVW1$ ({\em filled diamonds}), and $UVM2$ ({\em open diamonds}) bands,
compared with the flux evolution in the [$0.3-10$~keV] X-ray band ({\em open circles}).
Also shown are power-law fits ({\em dashed lines}) to the three X-ray LC segments , and
that to the observed $R$-band flux before the subtraction of the host galaxy flux ({\em solid line}).
\label{fig_LC} }

\plotone{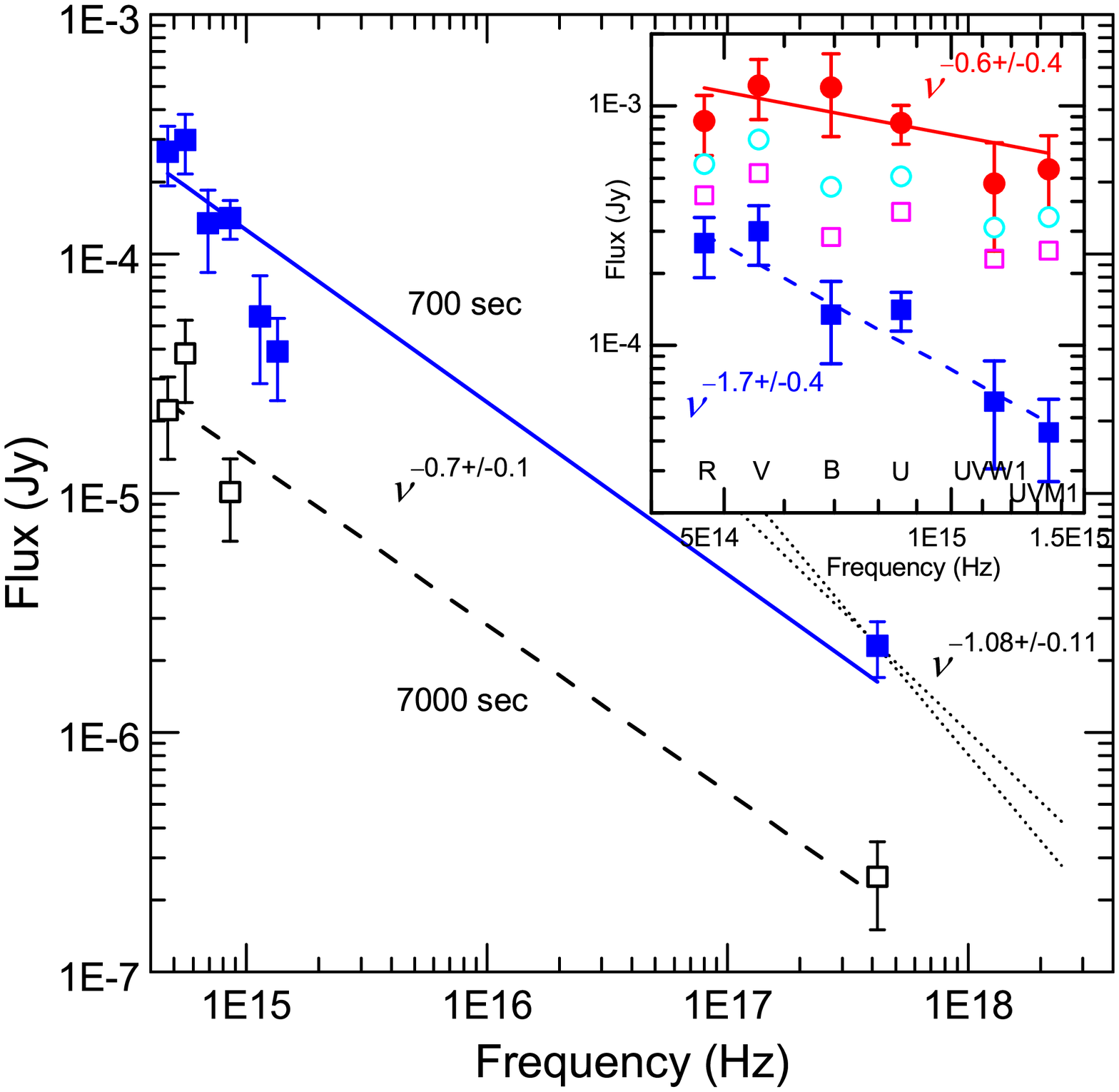}

\figcaption[f1.eps]{{\em Main} panel: Optical-to-X-ray SEDs of GRB 060912A afterglow
at 700~sec ({\em filled squares}) and 7,000~sec ({\em open squares}) after the BAT trigger,
with no correction of optical fluxes for H Lyman absorption and host dust extinction.
Also shown are power-law fits to the SEDs ({\em solid} and {\em dashed lines}) and
the fitting power-law X-ray spectra at 700~sec ({\em dotted lines}). {\em Inset}:
Optical SEDs of GRB 060912A afterglow at 700~sec, either uncorrected for host dust
extinction ({\em filled squares}) or corrected assuming the MW ({\em filled circles}),
LMC ({\em open circles}) and SMC ({\em open squares}) extinction law, respectively,
while H Lyman absorption and Galactic dust extinction has also been corrected.
Also shown are power-law fits to the MW case ({\em solid line)} and host extinction-uncorrected
case ({\em dashed line}).
\label{fig_sed} }

\begin{deluxetable}{lcccccccccr}
 \rotate
 \tabletypesize{\scriptsize}
 \tablecaption{Optical photometry of GRB 060912A afterglow}
 \tablewidth{0pt}
 \tablenum{1}
 \tablehead{
 \colhead{mean time (sec)\tablenotemark{a}}& \colhead{span (sec)\tablenotemark{b}}& \colhead{$R_C$}&  \colhead{$white$ \tablenotemark{c}}&
 \colhead{$V$}& \colhead{$B$}& \colhead{$U$}& \colhead{$UVW1$}& \colhead{$UVM2$}& \colhead{$UVOT$-$white$}& \colhead{telescope \tablenotemark{d}}
}
 \startdata
 9.90E1&  20 &        \nodata& 15.54$\pm0.08$&        \nodata&        \nodata&        \nodata&        \nodata&        \nodata&        \nodata& TNT \nl
 1.24E2&  20 &        \nodata& 15.76$\pm0.08$&        \nodata&        \nodata&        \nodata&        \nodata&        \nodata&        \nodata& TNT \nl
 1.50E2&  20 &        \nodata& 15.89$\pm0.08$&        \nodata&        \nodata&        \nodata&        \nodata&        \nodata&        \nodata& TNT \nl
 1.65E2&  100&        \nodata&        \nodata&        \nodata&        \nodata&        \nodata&\nodata&\nodata& 16.66$\pm0.03$&UVOT \tablenotemark{e}\nl
 1.87E2&  20 &        \nodata& 16.19$\pm0.08$&        \nodata&        \nodata&        \nodata&        \nodata&        \nodata&        \nodata& TNT \nl
 2.20E2&  20 &        \nodata& 16.38$\pm0.08$&        \nodata&        \nodata&        \nodata&        \nodata&        \nodata&        \nodata& TNT \nl
 2.47E2&  20 &        \nodata& 16.56$\pm0.08$&        \nodata&        \nodata&        \nodata&        \nodata&        \nodata&        \nodata& TNT \nl
 3.07E2&  80 &        \nodata& 16.88$\pm0.08$&        \nodata&        \nodata&        \nodata&        \nodata&        \nodata&        \nodata& TNT \nl
 4.17E2&  100&        \nodata& 17.30$\pm0.08$&        \nodata&        \nodata&        \nodata&        \nodata&        \nodata&        \nodata& TNT \nl
 4.20E2&  400&        \nodata&        \nodata& 17.57$\pm0.06$&        \nodata&        \nodata&        \nodata&        \nodata&        \nodata& UVOT\nl
 5.41E2&  100&        \nodata& 17.61$\pm0.08$&        \nodata&        \nodata&        \nodata&        \nodata&        \nodata&        \nodata& TNT \nl
 6.60E2&  20 &        \nodata&        \nodata&        \nodata&        \nodata&        \nodata& 18.33$\pm0.41$&        \nodata&        \nodata& UVOT\nl
 6.84E2&  20 &        \nodata&        \nodata&        \nodata&        \nodata& 17.61$\pm0.20$&        \nodata&        \nodata&        \nodata& UVOT\nl
 7.10E2&  170&        \nodata&        \nodata&        \nodata&        \nodata&        \nodata&        \nodata& 18.50$\pm0.43$&        \nodata& UVOT\nl
 7.17E2&  10 &        \nodata&        \nodata&        \nodata&        \nodata&        \nodata&        \nodata&        \nodata& 18.70$\pm0.22$& UVOT\nl
 7.60E2&  20 &        \nodata&        \nodata& 17.94$\pm0.32$&        \nodata&        \nodata&        \nodata&        \nodata&        \nodata& UVOT\nl
 7.77E2&  160&        \nodata&        \nodata&        \nodata& 18.85$\pm0.30$&        \nodata&        \nodata&        \nodata&        \nodata& UVOT\nl
 8.32E2&  20 &        \nodata&        \nodata&        \nodata&        \nodata& 18.17$\pm0.27$&        \nodata&        \nodata&        \nodata& UVOT\nl
 9.10E2&  100&        \nodata&        \nodata&        \nodata&        \nodata&        \nodata&        \nodata&        \nodata& 18.62$\pm0.07$& UVOT\nl
 1.10E3&  300& 18.31$\pm0.07$&        \nodata&        \nodata&        \nodata&        \nodata&        \nodata&        \nodata&        \nodata& TNT \nl
 1.16E3&  400&        \nodata&        \nodata& 18.72$\pm0.12$&        \nodata&        \nodata&        \nodata&        \nodata&        \nodata& UVOT\nl
 1.30E3& 1240&        \nodata&        \nodata&        \nodata&        \nodata&        \nodata&        \nodata& 18.98$\pm0.35$&        \nodata& UVOT\nl
 1.47E3&  10 &        \nodata&        \nodata&        \nodata&        \nodata&        \nodata&        \nodata&        \nodata& 18.70$\pm0.21$& UVOT\nl
 1.51E3&  180&        \nodata&        \nodata&        \nodata&        \nodata& 18.79$\pm0.28$&        \nodata&        \nodata&        \nodata& UVOT\nl
 1.56E3&  360& 18.73$\pm0.10$&        \nodata&        \nodata&        \nodata&        \nodata&        \nodata&        \nodata&        \nodata& TNT \nl
 1.61E3&  340&        \nodata&        \nodata&        \nodata& 19.98$\pm0.40$&        \nodata&        \nodata&        \nodata&        \nodata& UVOT\nl
 1.79E3&  10 &        \nodata&        \nodata&        \nodata&        \nodata&        \nodata&        \nodata&        \nodata& 19.24$\pm0.33$& UVOT\nl
 2.26E3&  400&        \nodata& 19.00$\pm0.08$&        \nodata&        \nodata&        \nodata&        \nodata&        \nodata&        \nodata& TNT \nl
 2.85E3&  480& 19.17$\pm0.10$&        \nodata&        \nodata&        \nodata&        \nodata&        \nodata&        \nodata&        \nodata& TNT \nl
 3.64E3&  600&        \nodata&        \nodata&        \nodata&20.07 (UL)\tablenotemark{f}&\nodata&    \nodata&        \nodata&        \nodata& TNT \nl
 3.80E3&  300&        \nodata&        \nodata&     19.40 (UL)&        \nodata&        \nodata&        \nodata&        \nodata&        \nodata& TNT \nl
 4.53E3& 1260&        \nodata&        \nodata&        \nodata&     20.60 (UL)&        \nodata&        \nodata&        \nodata&        \nodata& 60cm\nl
 5.22E3&  200&        \nodata&        \nodata&        \nodata&        \nodata& 19.84$\pm0.25$&        \nodata&        \nodata&        \nodata& UVOT\nl
 5.63E3&  200&        \nodata&        \nodata&        \nodata&        \nodata&        \nodata&        \nodata&        \nodata& 20.61$\pm0.19$& UVOT\nl
 6.14E3& 1530&        \nodata&        \nodata&        \nodata& 21.20$\pm0.40$&        \nodata&        \nodata&        \nodata&        \nodata& UVOT\nl
 6.75E3& 1630&        \nodata&        \nodata& 19.94$\pm0.31$&        \nodata&        \nodata&        \nodata&        \nodata&        \nodata& UVOT\nl
 7.06E3&  200&        \nodata&        \nodata&        \nodata&        \nodata&        \nodata&        \nodata&        \nodata& 20.86$\pm0.23$& UVOT\nl
 9.35E3& 8600& 20.78$\pm0.20$&        \nodata&        \nodata&        \nodata&        \nodata&        \nodata&        \nodata&        \nodata& TNT \nl
 1.13E4&  900&        \nodata&        \nodata&        \nodata&        \nodata&        \nodata&        \nodata& 21.15$\pm0.39$&        \nodata& UVOT\nl
 1.22E4&  900&        \nodata&        \nodata&        \nodata&        \nodata&        \nodata& 20.96$\pm0.29$&        \nodata&        \nodata& UVOT\nl
 1.28E4&  300&        \nodata&        \nodata&        \nodata&        \nodata& 20.52$\pm0.33$&        \nodata&        \nodata&        \nodata& UVOT\nl
 1.32E4&  440&        \nodata&        \nodata&        \nodata&        \nodata& 20.46$\pm0.31$&        \nodata&        \nodata&        \nodata& UVOT\nl
 1.61E4&  600& 21.06$\pm0.20$&        \nodata&        \nodata&        \nodata&        \nodata&        \nodata&        \nodata&        \nodata& 1.5m\nl
 1.87E4&  960& 21.55$\pm0.30$&        \nodata&        \nodata&        \nodata&        \nodata&        \nodata&        \nodata&        \nodata& 1.5m\nl
 1.99E4&  600& 21.07$\pm0.15$&        \nodata&        \nodata&        \nodata&        \nodata&        \nodata&        \nodata&        \nodata& 1.3m\nl
 2.06E4& 3600&        \nodata&        \nodata&        \nodata& 22.44$\pm0.30$&        \nodata&        \nodata&        \nodata&        \nodata& 1.5m\nl
 2.14E4& 1800& 21.10$\pm0.15$&        \nodata&        \nodata&        \nodata&        \nodata&        \nodata&        \nodata&        \nodata& 1.3m\nl
 2.28E4& 2160& 21.57$\pm0.20$&        \nodata&        \nodata&        \nodata&        \nodata&        \nodata&        \nodata&        \nodata& 1.5m\nl
 2.39E4& 1800& 21.36$\pm0.20$&        \nodata&        \nodata&        \nodata&        \nodata&        \nodata&        \nodata&        \nodata& 1.3m\nl
 \enddata
 \tablenotetext{a}{Middle time for a single exposure, or exposure-time-weighted mean time for multiple exposures.}
 \tablenotetext{b}{Time span from the beginning of first exposure to the end of last exposure.}
 \tablenotetext{c}{$R$-band equivalent; TNT unfiltered photometry calibrated using $R$-band reference stars.}
 \tablenotetext{d}{The telescopes are the 80-cm TNT and 60-cm of Xinglong Obs., 1.3-m of Skinakas Obs., 1.5-m of Maidanak Obs., and UVOT of $Swift$.}
 \tablenotetext{e}{The UVOT photometry is derived using data extracted from http://swift.gsfc.nasa.gov/docs/swift/results/uvot\_grb/cat/.}
 \tablenotetext{f}{UL means an upper limit only.}
\end{deluxetable}

\begin{deluxetable}{ccccccc}
 \tabletypesize{\scriptsize}
 \tablecaption{Host galaxy-subtracted magnitudes \tablenotemark{a} at different epoches}
 \tablewidth{0pt}
 \tablenum{2}
 \tablehead{
 \colhead{epoch (sec)} &\colhead{$R_C$} & \colhead{$V$} & \colhead{$B$} & \colhead{$U$} & \colhead{$UVW1$} & \colhead{$UVM2$}
}
 \startdata
                  700  & 17.78$\pm0.3$ & 17.87$\pm0.3$ & 18.79$\pm0.4$ & 17.68$\pm0.2$ &   18.46$\pm0.5$  &  18.55$\pm0.4$ \nl
                 7,000 & 20.48$\pm0.4$ & 20.10$\pm0.4$ &    \nodata    & 20.54$\pm0.4$ &      \nodata     &     \nodata    \nl
 \enddata
 \tablenotetext{a}{The small Galactic reddening of $E(B-V)=0.05$ has not been corrected for.}
\end{deluxetable}

\end{document}